\input{aipcheck.tex}

\newcommand\sps{\space\space\space\space}
\def\selectedoptions{final}

\documentclass[
   \selectedoptions
  ]
  {aipproc}

\layoutstyle{6x9}

\SetInternalRegister\hbadness{8000} 

\newcommand\doingARLO[2][]{%
  \ifx\mmref\undefined #1\else #2\fi
}
  
\usepackage{natbib}
\begin{document}

\title 
      [Magnetic field stability of Vega]
      {Long-term magnetic field stability of Vega}


\author{D. Alina}{
  address={Universit\'e de Toulouse, UPS-OMP, Institut de Recherche en Astrophysique et Plan\'etologie, Toulouse, France},
  altaddress={CNRS, Institut de Recherche en Astrophysique et Plan\'etologie, 14 Avenue Edouard Belin, F-31400 Toulouse, France}
}

\iftrue
\author{P. Petit}{
  address={Universit\'e de Toulouse, UPS-OMP, Institut de Recherche en Astrophysique et Plan\'etologie, Toulouse, France},
  altaddress={CNRS, Institut de Recherche en Astrophysique et Plan\'etologie, 14 Avenue Edouard Belin, F-31400 Toulouse, France}
}

\author{F. Ligni\`eres}{
  address={Universit\'e de Toulouse, UPS-OMP, Institut de Recherche en Astrophysique et Plan\'etologie, Toulouse, France},
  altaddress={CNRS, Institut de Recherche en Astrophysique et Plan\'etologie, 14 Avenue Edouard Belin, F-31400 Toulouse, France}
}

\author{G.A. Wade}{
  address={Department of Physics, Royal Military College of Canada, PO Box 17000, Station Forces, Kingston, Ontario, Canada},
}

\author{R. Fares}{
  address={School of Physics and Astronomy, University of St Andrews, St Andrews, Scotland KY16 9SS},
}

\author{M. Auri\`ere}{
  address={Universit\'e de Toulouse, UPS-OMP, Institut de Recherche en Astrophysique et Plan\'etologie, Toulouse, France},
  altaddress={CNRS, Institut de Recherche en Astrophysique et Plan\'etologie, 14 Avenue Edouard Belin, F-31400 Toulouse, France}
}

\author{T. B\"ohm}{
  address={Universit\'e de Toulouse, UPS-OMP, Institut de Recherche en Astrophysique et Plan\'etologie, Toulouse, France},
  altaddress={CNRS, Institut de Recherche en Astrophysique et Plan\'etologie, 14 Avenue Edouard Belin, F-31400 Toulouse, France}
}

\author{H. Carfantan}{
  address={Universit\'e de Toulouse, UPS-OMP, Institut de Recherche en Astrophysique et Plan\'etologie, Toulouse, France},
  altaddress={CNRS, Institut de Recherche en Astrophysique et Plan\'etologie, 14 Avenue Edouard Belin, F-31400 Toulouse, France}
}
\fi

\copyrightyear  {2001}

\begin{abstract}
We present new spectropolarimetric observations of the normal A-type star Vega, obtained during the summer of 2010 with NARVAL at T\'elescope Bernard Lyot (Pic du Midi Observatory). This new time-series is constituted of 615 spectra collected over 6 different nights. We use the Least-Square-Deconvolution technique to compute, from each spectrum, a mean line profile with a signal-to-noise ratio close to 20,000. After averaging all 615 polarized observations, we detect a circularly polarized Zeeman signature consistent in shape and amplitude with the signatures previously reported from our observations of 2008 and 2009. The surface magnetic geometry of the star, reconstructed using the technique of Zeeman-Doppler Imaging, agrees with the maps obtained in 2008 and 2009, showing that most recognizable features of the photospheric field of Vega are only weakly distorted by large-scale surface flows (differential rotation or meridional circulation). 
\end{abstract}

\date{\today}

\maketitle

\section{Introduction}

The detection of Zeeman signatures formed in the photosphere of the normal, rapidly-rotating A-type star Vega was the first detection of a sub-gauss magnetic field on a main-sequence star of intermediate mass that does not belong to the class of Ap/Bp stars \citep{lignieres2009}. This discovery was later followed by the detection of polarized signatures on the Am star Sirius A \citep{petit2011}, suggesting that Vega-like magnetism may be a widespread property of stars in this mass domain. Confronted with what might be a new type of stellar magetism, we aim at gathering more observational clues about the origin of these weak surface fields and providing theoreticians with observational constraints that can guide them towards identifying and modelling the physical ingredients at the origin of Vega-like magnetism. With this in mind, we are now engaged in a long-term monitoring of Vega, in order to investigate the stability of its magnetic geometry, and to determine whether its magnetic variability, if any, can be related to specific large-scale surface flows, like latitudinal differential rotation or meridional motions.  

\section{Spectropolarimetric observations}

We used the NARVAL spectropolarimeter \citep{auriere2003} to collect a new time-series of observations of Vega during the summer of 2010. The data set consists in a total of 615 spectra obtained over 6 nights (5 consecutive nights in July, complemented by one isolated night in August). The instrumental setup allows us to simultaneously record spectra in light intensity (Stokes I) and circular polarization (Stokes V) in a nearly continuous wavelength domain between 370 and 1,000~nm, at a spectral resolution of $\approx 65,000$. 

\section{Zeeman signatures}

To reach the extremely high signal-to-noise ratio required to detect the tiny Zeeman signatures of Vega, the Least-Squares Deconvolution (LSD) cross-correlation technique \citep{donati1997} was applied to all spectra using a list of about 1,100 photospheric lines, producing from each spectrum a mean line profile with signal-to-noise-ratio (S/N) close to 20,000.  In order to further reduce the noise, we computed a temporal average of all 615 LSD profiles, to finally detect a Stokes V signature close to the central radial velocity of the LSD line profile. The shape and amplitude of the signature do not exhibit any measurable difference compared to the signatures derived from the campaigns of 2008 and 2009 \citep{lignieres2009, petit2010}. Accordingly, the strength of the longitudinal field component derived from the July 2010 data, calculated using the center of gravity technique \citep{rees1979}, yields $B_l = -0.3\pm 0.2$ G, in close agreement with earlier estimates. 

\section{Period search}

\begin{figure}
  \reflectbox{\includegraphics[height=.4\textheight]{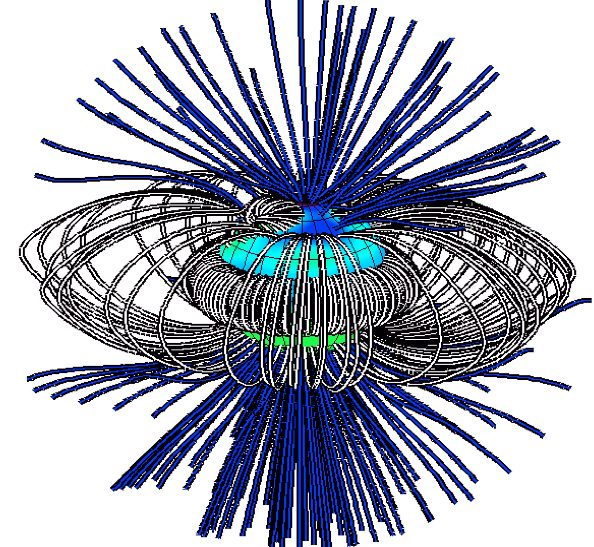}}
  \caption{Extrapolation of the potential surface magnetic field above stellar surface (source surface assumed to be at 2.5 stellar radii). Closed field lines in white, open ones in blue.
}
\label{fig:corona}
\end{figure}

A search for rotational modulation of the polarized signatures was conducted, by applying a simple least squares sine fit period search to each bin in radial velocity, for a period range between 0.4 and 1 day, over the time-series of 500 Stokes V LSD profiles  obtained in July. When the search is restricted to the spectral bins inside the line profile, a $\chi^2$ minimum is obtained for a period of $0.678^{+0.036}_{-0.029}$~d, while no such deep $\chi^2$ minimum can be identified for velocity bins outside of the line profile. Following the method of \cite{petit2010}, an indirect period search was also carried out using the Zeeman Doppler Imaging technique (ZDI, \cite{donati1997,donati2006}). Using this second approach, three $\chi^2$ minima of similar depths are recovered between 0.4~d and 1~d, one of them being located at 0.68~d, in good agreement with the direct period search.

\section{Magnetic topology}

In 2010, the reconstructed magnetic field is mostly poloidal, with 83\% of the magnetic energy in the poloidal component, and is mostly axisymmetric (with 59.73\% of the magnetic energy in modes with $m = 0$). The magnetic topology is complex, with only 9\% of the magnetic energy in the dipole, 5\% in the quadrupole and 6\% in the octopole; most of the magnetic energy is concentrated in higher-order  spherical harmonics components ($\ell > 3$). The large-scale magnetospheric structure of Vega was extrapolated from the potential component of the surface magnetic field, using the Zeeman-Doppler Imaging magnetic map as the photospheric  boundary conditions \citep{jardine2002}, assuming a source surface at 2.5 stellar radii (Fig. \ref{fig:corona}). 

\section{Discussion}

The rotation period directly derived from the Stokes V signatures is close to, though slightly shorter than, the estimates of \cite{takeda2008} and \cite{petit2010}, who find, instead, a period close to 0.73~d (with error bars of $\approx 0.01$~d for \cite{petit2010}). Our value of the period is more difficult to reconcile with the shorter estimate of \cite{aufdenberg2006}, based on long baseline interferometry. By applying the same sine fit method to the Stokes V profiles collected in 2008 and 2009, we derive period estimates that agree well with the measurement obtained for 2010. Further investigations are needed to interpret the origin of the differences between the outcomes of the various period searches.
 
The magnetic geometry reconstructed from the 2010 data set is consistent with maps previously derived from July 2008 and September 2009 observations \citep{petit2010}, revealing the stability of Vega's weak magnetic field over the 3 years of observation, although the low inclination angle of Vega is probably not adequate for a fine tracking of variability in the equatorial region (which is seen close to the stellar limb). The observed stability suggests that large-scale surface flows, if any, may distort the magnetic topology much more slowly than those of cooler stars possessing convective outer layers (e.g. \cite{barnes2005}). A longer temporal monitoring may help unveil very slow plasma motions that may progressively shape the surface magnetic field. Such observations may help select between various options at our disposal to explain the origin of Vega's magnetic field \citep{auriere2007,rudiger2011,moss2001}. 

\doingARLO[\bibliographystyle{aipproc}]
          {\ifthenelse{\equal{\AIPcitestyleselect}{num}}
             {\bibliographystyle{arlonum}}
             {\bibliographystyle{arlobib}}
          }
\bibliography{alina_d}

\end{document}